\documentclass[12pt, preprint]{aastex}
\usepackage{natbib}
\usepackage{lineno}

\shorttitle{Tycho with the \emph{Fermi}-LAT}
\shortauthors{Giordano, Naumann-Godo et al.}

\begin{document}

\title{\emph{Fermi}-LAT Detection of the Young  SuperNova Remnant Tycho}

\author{
F.~Giordano\altaffilmark{1,2,*},
M.~Naumann-Godo\altaffilmark{3,**},
J.~Ballet\altaffilmark{3},
K.~Bechtol\altaffilmark{4},
S.~Funk\altaffilmark{4},
J.~Lande\altaffilmark{4},
M.~N.~Mazziotta\altaffilmark{2},
S.~Rain\`o\altaffilmark{2},
T.~Tanaka\altaffilmark{4},
O.~Tibolla\altaffilmark{5},
Y.~Uchiyama\altaffilmark{4}
}
\altaffiltext{1}{Dipartimento di Fisica ``M. Merlin" dell'Universit\`a e del Politecnico di Bari, I-70126 Bari, Italy}
\altaffiltext{2}{Istituto Nazionale di Fisica Nucleare, Sezione di Bari, 70126 Bari, Italy}

\altaffiltext{3}{Laboratoire AIM, CEA-IRFU/CNRS/Universit\'e Paris Diderot, Service d'Astrophysique, CEA Saclay, 91191 Gif sur Yvette, France}

\altaffiltext{4}{W. W. Hansen Experimental Physics Laboratory, Kavli Institute for Particle Astrophysics and Cosmology, Department of Physics and SLAC National Accelerator Laboratory,
Stanford University, Stanford, CA 94305, USA}
\altaffiltext{5}{Institut f\"ur Theoretische Physik and Astrophysik, Universit\"at W\"urzburg, D-97074 W\"urzburg, Germany}
\altaffiltext{*}{email: francesco.giordano@ba.infn.it}
\altaffiltext{**}{email: Melitta.Naumann-Godo@cea.fr}

\begin{abstract}
After almost three years of data taking in sky survey mode, the \emph{Fermi}-LAT has detected $\gamma$-ray emission toward the Tycho's Supernova Remnant (SNR). The Tycho SNR is among the
youngest remnants in the Galaxy, originating from
a Type Ia Supernova in AD 1572.
The $\gamma$-ray integral flux from 400~MeV up to 100~GeV has been measured  to be
(3.5$\pm$1.1$_{stat}\pm$0.7$_{syst}$)$\times$10$^{-9}$ cm$^{-2}$s$^{-1}$ with a   photon index of 2.3$\pm$0.2$_{stat}\pm$0.1$_{syst}$.

A simple model consistent with TeV, X-ray and radio data is sufficient to explain
the observed emission as originating from $\pi^0$-decays as a result of cosmic-ray acceleration and interaction with the ambient medium.

\end{abstract}


\section{Introduction}

Tycho's SNR (SN 1572 or SNR120.1+01.4) is classified as a Type Ia (thermonuclear explosion of a
white dwarf) based on observations of the light-echo spectrum \citep{krause}.
Its expansion has been observed in the radio \citep[VLA,][]{vla}
and X-rays \citep[Chandra,][]{katsuda_2010}.
The average expansion parameter is $\nu$ = 0.47 (radio)
to 0.52 (X-rays) and shows strong azimuthal variations.
Along 3/4 of the rim it is consistent with $\nu$ = 4/7, expected from
a reverse shock developing into $r^{-7}$
ejecta \citep{chevalier82}, while toward the east it is much lower, probably
due to recent deceleration in a higher density medium.
The shock speed is approximately
1400 $D_{\rm kpc}$ km s$^{-1}$ except toward the east.

Tycho's distance is not very well constrained.
H~{\sc I} absorption studies \citep{tian} favor a distance on the near side of the Perseus arm (2.5 to 3 kpc).
On the other hand
\citet{suzaku} measured the shocked ejecta velocity to be about 4700 km s$^{-1}$
for Si, S and Ar. They concluded that the distance is 3 to 5 kpc
on consideration of the measured proper motion.

The radio flux \citep{kothes_2006} is 40.4 Jy at 1.4 GHz with a spectral index of 0.65.
The overall X-ray spectrum \citep{xmm} is dominated by very strong line
features of Si, S and Fe arising in the shocked ejecta.
However the major fraction (60\%) of the X-ray continuum emission comes from
non-thermal synchrotron rather than thermal bremsstrahlung \citep{chandra2},
and non-imaging instruments indicate that it extends to more than 10 keV
\citep{fink,petre}.
Recently the VERITAS ground based telescope reports a TeV detection with a flux of 0.9$\%$ of the steady Crab Nebula \citep{veritas}.

Several arguments point toward a cosmic-ray-modified shock in Tycho.
\cite{chandra2} have argued that the contact discontinuity between the
ejecta and the shocked ambient medium is too close to the blast wave
for a compression factor of 4.
The other important hint comes from the narrow width of the X-ray
synchrotron rims, which is probably due to fast cooling of the accelerated
electrons behind the blast wave, and requires the magnetic field
to be amplified by accelerated particles to 200 $\mu$G \citep{chandra1}.
Recent optical observations of a fast shock in Tycho have even
provided the first observational indications of a cosmic-ray precursor
\citep{lee2010}.

In this letter we report the GeV detection of Tycho's SNR with the \emph{Fermi}-LAT.

\section{Observations and Data Reduction}

The {\em Fermi $\gamma$-ray Space Telescope}, launched in June 2008,
has operated almost exclusively in sky-survey mode since 2008 August 4.
The principal \emph{Fermi} scientific instrument, the Large Area Telescope (LAT)  has a wide field-of-view  of 2.4~sr,
which allows it to observe the whole sky every $\sim 3$~hr (2 orbits), a large effective area of $\sim 8000~{\rm cm}^2$ (on axis at 1~GeV)
and a point-spread function (PSF) narrower than $1.0^\circ$ (for 68\% containment)
at 1~GeV.
For a detailed description of the instrument see \cite{atwood}.

We have selected a data set of 34 months  in a square region of interest (RoI) of 20$^{\circ}$ side length, centered on the position of the remnant,
selecting events  from 400~MeV to 100~GeV in the
$P6\_DIFFUSE$ class\footnote{$http://www-glast.slac.stanford.edu/software/IS/$}.
The choice of selecting events above 400~MeV is motivated by the broad PSF at low energy ($\approx$5$\fdg$ at 100~MeV). Moreover  the SNR is only 1.4$\degr$
 from the Galactic plane, a
region mainly dominated by low-energy photons from the
Galactic diffuse emission, which makes the low-energy determination particularly difficult and subject to systematics.

In addition, to limit contamination from photons
produced by cosmic-ray interactions in the upper atmosphere,
a 105$^{\circ}$ cut on the Earth zenith angle has been applied \citep{earth_albedo}.

\section{Analysis and Results}
\begin{figure*}[bhpt]
\epsscale{0.4}
\plotone{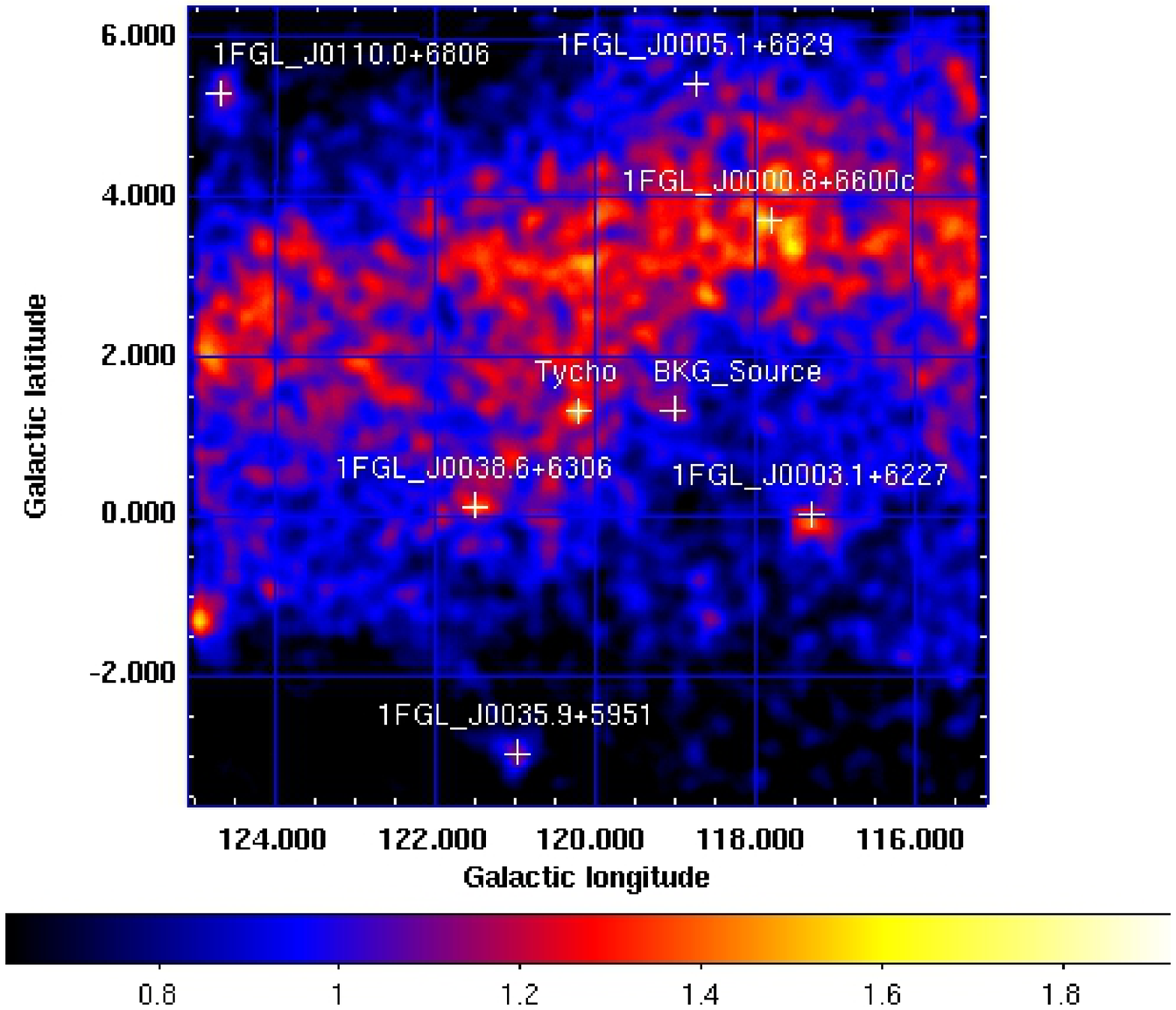}
\epsscale{0.41}
\plotone{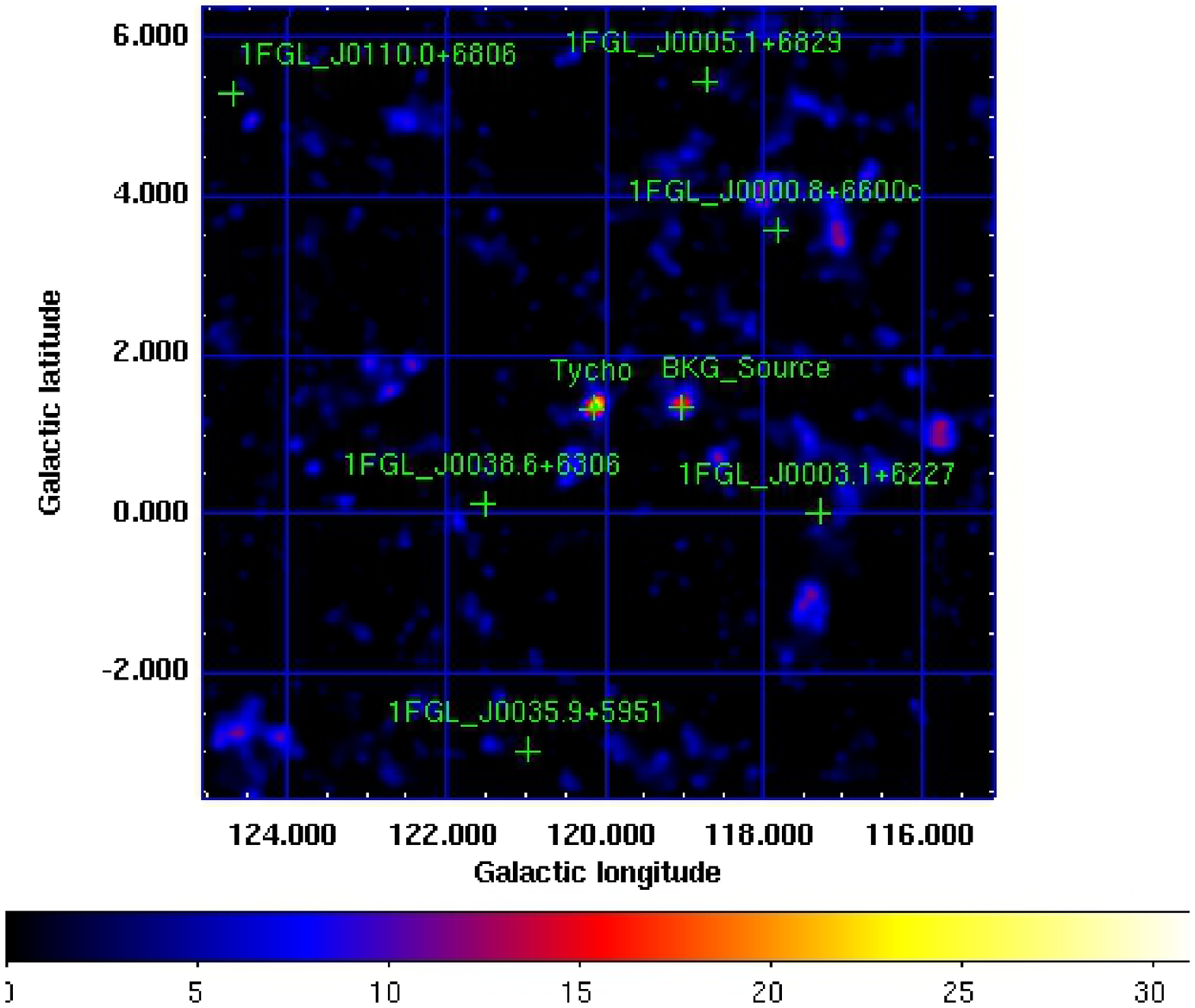}
\caption{left: \emph{Fermi}-LAT count map of the region of 10$\times$10$\degr$ for photons above 1GeV. The crosses indicate the pointlike sources included in the model for the fit.
A gaussian smoothing of 0.3$^{\circ}$ has been applied. right: Map of the likelihood TS (i.e., map of the TS values obtained for a test source that has been stepped
through a grid of positions, with the likelihood re-optimized at each position) of the same region for photons above 1GeV. All contributions
from known sources and diffuse emission have been included in the source model and so are not apparent here
except Tycho, which corresponds to the bright spot (TS$>$25) in the center of the image.
\label{Map} }
\end{figure*}

The $\gamma$-ray spectrum from a point-like source coincident with the Tycho SNR has been obtained using the Fermi Science Tool
$gtlike$ publicly available from the  Fermi Science Support Center (FSSC)\footnote{$http://fermi.gsfc.nasa.gov/ssc/$}.
 $gtlike$   has been used in two ways in binned mode:
the first method consists in fitting
all sources in the RoI assuming a  simple power law model
in the selected energy interval;
both the integral flux and the photon index of each source within 5 $\degr$ from Tycho's SNR are kept free
in the fit.
Only the pulsar PSRJ0007+7303 in the CTA1 SNR  has been fitted with a power law with an exponential cut-off.
  Moreover, all point-like sources
included in the 1FGL catalog \citep{1fgl} and located within the selected RoI,  have been accounted for (Fig.~\ref{Map} left).
The  diffuse emission, including the emission from our Galaxy and an isotropic component,  is modeled according to the rescaled gll\_iem\_v02 template  optimized for the Instrument Response
Function (IRF) $P6V11$.
Figure~\ref{Map} right shows the map of the likelihood Test Statistic (TS)\footnote{The Test Statistic is defined as TS = 2(log(L$_1$)-log(L$_0$)) with L$_0$ the
likelihood of the Null-hypothesis model (no point source present) as compared to the likelihood of
a competitive model, L$_1$.}  for E$>$1GeV of a region 10$\times$10 $\degr$ around Tycho. The map has been obtained including all the contributions from the backgrounds (point sources
and diffuse)
except Tycho: a bright spot ($TS>25$) is visible in the center of the map, which corresponds to the \emph{Fermi}-LAT detection of Tycho. Another peak,  about one degree away from Tycho,
is visible and has been added to the model.

\begin{figure}[htb]
\begin{center}
\includegraphics[scale=.4]{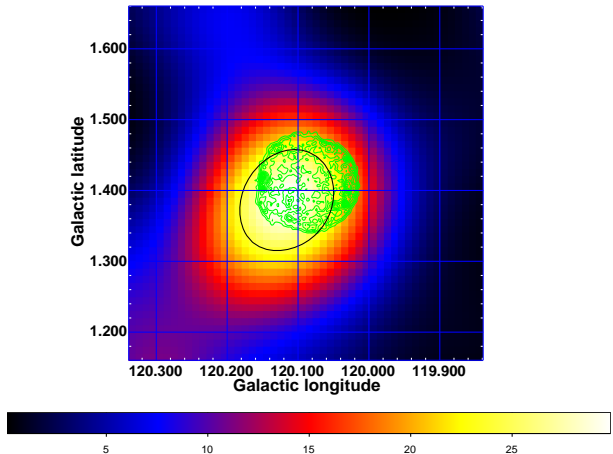}
\end{center}
\caption{\emph{Fermi}-LAT TS map zoomed in.
The green contours are  4.5~keV - 5.8~keV continuum band from XMM-Newton \citep{xmm} and the black line denotes the 95\% confidence area for the \emph{Fermi}-LAT position.}
\label{TS_1GeV}
\end{figure}

The position of the point-like GeV source is
obtained using the tool $gtfindsrc$ selecting events above 1GeV.
The results of the localization based on a maximum likelihood fit are $R.A.=6\fdg45$ and  $Dec=64\fdg12$ (J2000) with a 68\% error radius of 0.03$^{\circ}$. Systematic uncertainties are estimated
to be about 0.006$^{\circ}$ as stated in \cite{1fgl}; therefore we cannot  be more precise in localizing   the emission within the SNR.
The most likely position for the $\gamma$-ray emission above 1~GeV is shown in Figure~\ref{TS_1GeV} as the 95\% confidence contour, which is in agreement with the radio and X-ray template.
Moreover, a study comparing the difference in TS between a point-like source at the above mentioned position and the XMM-Newton template  yields
a small TS-difference of 3.3. Since this difference is not significant, given the two degrees of freedom ($dof$) allowed for the position, this confirms that the localization of
Tycho's remnant in X-rays is fully compatible with the \emph{Fermi}-LAT data,
while no further information  on size and   morphology can be inferred.
Concerning the spectral properties, the overall fit from 400~MeV to 100~GeV yields a TS of 33 for Tycho, which corresponds to a detection of about 5$\sigma$ for point-like sources with 4~$dof$.
The measured integral flux is (3.5$\pm$1.1$_{stat}\pm$0.7$_{syst}$)$\times$10$^{-9}$ cm$^{-2}$s$^{-1}$ with a photon index of 2.3$\pm$0.2$_{stat}\pm$0.1$_{syst}$.
The additional background source
at R.A.,Dec = ($3\fdg94$, $63\fdg94$) has a comparable overall flux of (3.3$\pm$2.0$_{stat}$)$\times$10$^{-9}$ cm$^{-2}$s$^{-1}$
with a softer photon index of 2.4$\pm$0.2$_{stat}\pm$0.1$_{syst}$.
Its significance
is 5$\sigma$.

\begin{figure}[htb]
\begin{center}
\includegraphics[scale=.4]{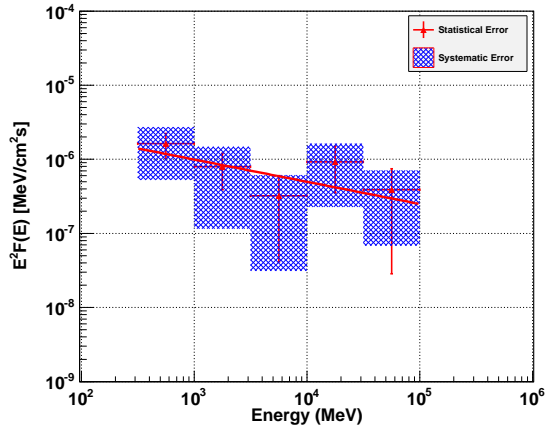}
\end{center}
\caption{Spectrum obtained by evaluating the flux in separate energy bins. The shadowed regions indicate the uncertainties due to systematics in the effective area and in the accuracy of the modeling of the Galactic diffuse emission. The solid line reproduces  the best fit  for a power-law model.}
\label{fig:sed}       
\end{figure}

The second method consists of dividing the entire energy range in logarithmically spaced energy bins from 400~MeV to 100~GeV and in fitting all point sources and the scale
factors for the diffuse components in
each bin.
The fitting procedure assumes  simple power law models for all the sources in the RoI and in each energy bin with a  photon index arbitrarily fixed at 2. The results
of  this band-by-band fitting procedure are shown in Figure~\ref{fig:sed} for Tycho.
In this figure the data points are drawn with statistical uncertainties only; the systematics have been indicated with dashed boxes, calculated making the same fit with different IRFs
(the Monte Carlo-based P6V3 and the in-flight corrected P6V11) and different models of the Galactic and isotropic diffuse emission optimized for each IRFs.

\section{Discussion}

\subsection{Association}
A point-like GeV $\gamma$-ray source is found in the direction of Tycho's SNR
based on more than 2.5 years of observation with the \emph{Fermi}-LAT.
The positional coincidence between the \emph{Fermi}-LAT $\gamma$-ray
emission and Tycho's SNR  suggests that the $\gamma$-ray
emission is produced by shock-accelerated particles within the remnant.
We have also considered the possibility to have a $\gamma$-ray emission from a pulsar, that is the class of most numerous Galactic objects.
We have excluded this possibility   considering
that  Tycho's SNR is known to be
the result of a type Ia explosion \citep{ruiz-lapuente, krause}
and that the closest radio pulsar in the line of sight, PSR J0026+6320  (about 0.8$^{\circ}$  from Tycho),
has a   dispersion measure distance of about 13.6~kpc with a  spin-down power of approximately 2$\times$10$^{32}$erg s$^{-1}$.
Moreover, the $\gamma$-ray spectrum measured with the \emph{Fermi}-LAT
does not resemble the spectrum of a typical pulsar usually featuring
an exponential cutoff \citep{pulsarcat}.
Also the chance probability of finding an AGN (the most populated class of LAT sources) within the 95\% confidence contour has been investigated.
The surface density of an AGN with an integral flux equivalent
to that measured for Tycho or even greater is less than 8$\times$10$^{-3}$/deg$^2$ \citep{lognlogs}; therefore
 the chance probability of finding one within 5$\arcmin$  is less than 10$^{-4}$.

\subsection{General constraints}

We consider the GeV $\gamma$-ray production from the bulk of the SNR
(not the special eastern region).
\citet{chandra1} have analyzed the blast wave of Tycho in the framework
of a model that accounts for the dynamical effect of diffusive shock
acceleration and is broadly consistent with the location of the ejecta,
that requires the shock compression at the blast wave to be  $\sim$6.
That model assumes a self-similar evolution driven by ejecta distributed
as $r^{-7}$, consistent with
the expansion measurements of \citet{vla} and \citet{katsuda_2010}.
\citet{truelove}
have shown that the blast wave radius follows the self-similar solution
up to twice the time when the reverse shock reaches the central plateau.
So that solution is valid for the densities considered below up to a distance
of 4 kpc or so.

As discussed in the \S Introduction, the main uncertainty
is the distance to the source. Therefore we need to extract the explicit
dependence of the observational constraints on distance (expressed in kpc)
$D_{\rm kpc}$.
The explosion energy $E_{51}$ (in units of $10^{51}$ erg) and ambient density
$n_H$ are not known precisely either.
In order to reproduce the angular size of Tycho at the present time
($4.26'$ in the west), \cite{chandra1} obtain a relation between density and distance, assuming
$E_{51} = 1$. This can be easily generalized to any explosion
energy, resulting in
\begin{equation}
\label{angsize}
n_H = 388 D_{\rm kpc}^{-7} E_{51}^2 {\rm cm}^{-3}
\end{equation}

The absence of a clear detection of thermal X-ray emission from the shocked
ambient gas led \cite{chandra1} to a constraint on density, namely
$n_H < 0.3$ cm$^{-3}$ at a distance of 2.8 kpc.
That constraint was based on X-ray brightness and it does not depend
very much on temperature, so it depends little on $E_{51}$.
Since X-ray brightness is proportional to
$\int n^2 dl$ along the line of sight, the main dependence is as
$n_H^2 D_{\rm kpc}$. Therefore the constraint can be generalized to
$n_H  < 0.5 D_{\rm kpc}^{-0.5}$cm$^{-3}$.

We also account for the fact that the total available energy
at the blast wave $E_{51}^{\rm amb}$ is not the full explosion energy,
as some of it is still locked up in the ejecta.
In the self-similar phase, it can be estimated as
\begin{equation}
\label{shockedenergy}
E_{51}^{\rm amb} = 0.24 n_H^{1/2} D_{\rm kpc}^{3/2} E_{51}
\end{equation}
Extrapolating that formula beyond the time at which the reverse shock enters
the central plateau is reasonable. The transfer of energy to the shocked
ambient gas continues, coming then from the shocked ejecta.
$E_{51}^{\rm amb}/E_{51}$ would be 0.93 at the end of the validity
of the self-similar phase defined above, so this approximation connects
smoothly to the Sedov phase in which $E_{51}^{\rm amb} = E_{51}$.

For either bremsstrahlung or $\pi^0$ decay which use the same target gas,
the predicted $\gamma$-ray flux will be
\begin{equation}
\label{gammaflux}
F_\gamma \propto f_{\rm CR} E_{51}^{\rm amb} 3 n_H D_{\rm kpc}^{-2}
         \propto 0.72 f_{\rm CR} E_{51} n_H^{3/2} D_{\rm kpc}^{-1/2}
\end{equation}
where we have used Eq. \ref{shockedenergy}.

Eqs. \ref{angsize} and \ref{gammaflux} define, as a function of distance,
a family of solutions in which $n_0$ decreases as $D_{\rm kpc}^{-3/2}$
and $E_{51}$ increases as $D_{\rm kpc}^{11/4}$.

\subsection{$\gamma$-ray emission}
There are three radiation processes potentially
responsible for the GeV $\gamma$ rays from (or in the direct
vicinity of) Tycho's SNR:
inverse-Compton (IC) scattering on the cosmic microwave background (CMB) by
relativistic electrons;
nonthermal bremsstrahlung by relativistic electrons;
and $\pi^0$-decay $\gamma$ rays resulting mainly
from inelastic collisions between relativistic
protons and ambient gas nuclei \citep{gaisser_1998}.

Two different cases are considered in the following:
in the first case, called the "nearby" scenario,
the total energy output of the supernova will be
fixed to a standard value of $10^{51}$~erg, which
results in a distance of 2.78~kpc assuming the maximally allowed value for the ambient density of 0.3~cm$^{-3}$.
In the second case,
we place the remnant at a distance of 3.5~kpc and calculate the supernova energy and its ambient density to be 2$\times$10$^{51}$~erg
and 0.24~cm$^{-3}$ respectively. This constitutes the "far" scenario.

In both cases, the synchrotron flux is constrained by the radio and X-ray data.
These data imply that a population of shock-accelerated electrons described by a power-law spectrum with a spectral index of
2.2--2.3 and a cut-off energy of 6-7 TeV is the origin of the observed synchrotron emission
if we assume the downstream magnetic field to be $B_{\rm d} \sim 215 \mu$G as inferred by
X-ray measurements  \citep{chandra1}.
Since the magnetic field strength is the only parameter
to determine the strength of the IC flux produced by the above mentioned population of electrons on the CMB radiation and IR photon fields,
the IC flux at $\sim 1$ GeV is inevitably far below the observed value.

Bremsstrahlung is the only way to possibly fit the data in a leptonic model.
The flux of bremsstrahlung $\gamma$ rays scales with the electron population and $n_{\rm H}$.
Since the electron population cannot be increased too much, otherwise the IC contribution would quickly exceed the TeV data, the ambient density needs to be increased and the downstream
magnetic field has to be decreased to $65\ \mu$G. In the most favorable case of the nearby scenario, the required gas density would be $n_{\rm H}=9.7$~cm$^{-3}$
(using an effective density twice larger behind the shock, as prescribed by a Sedov model). This value exceeds the gas density allowed by X-ray measurements by a factor of 30  \citep{chandra1}.
 Moreover, given the extremely high gas density, the energy in the accelerated hadrons needs to be severely reduced to at least 1.5$\times 10^{48}$~erg which would correspond to an extremely
 high electron to proton ratio ($K_{ep}=0.1$). In this case the SNR has to be in the Sedov phase and thus the supernova energy would be $4.4\times 10^{51}$~erg, which is exceptionally high for
  a type Ia supernova. For all these reasons the bremsstrahlung and IC channel are very unlikely to account for the \emph{Fermi}-LAT measurement.

\begin{figure}
\begin{center}
\includegraphics[width=0.95\linewidth]{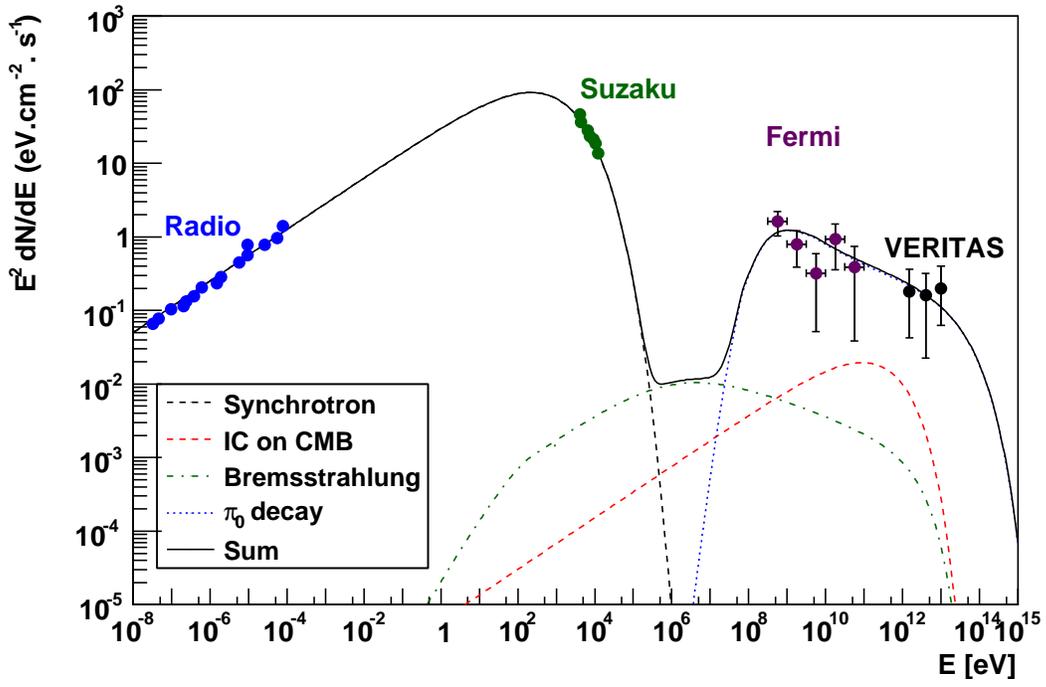}
\caption{Broadband SED model of Tycho's SNR for the far scenario.}
\label{fig:mwsed}
\end{center}
\end{figure}

On the other hand, the expected $\gamma$-ray spectrum of hadronic origin can be calculated
on the assumption that efficient proton acceleration is taking place at
the forward shock in Tycho's SNR, whereby the protons acquire the same power-law spectrum with
the same spectral index as the aforementioned electron spectrum. Only the cut-off energy is much higher
as protons do not suffer from synchrotron losses as much as electrons do.
Since the VERITAS spectral measurements \citep{veritas} do not indicate any cut-off, we estimate the maximum proton energy by equating the acceleration time (assuming Bohm diffusion)
with the age of Tycho. For $B_{\rm d}$ = 215 $\mu$G, this results in an energy
break for protons of $E_{p,max}$ = 44 $D_{\rm kpc}$$^2$ TeV \citep{pcut}.

The intensity of this emission depends on the total energy in the accelerated protons (and other ions) as well as on the density of the ambient medium $n_{\rm H}$. Even though the
density is expected to be around 6 $n_{\rm H}$ just behind the shock front, the average density seen by the shock-accelerated protons over the full emission zone between the blast wave and
the ejecta is only around 3 $n_{\rm H}$ in the self-similar model.

Furthermore, the $\gamma$-ray emission has been computed assuming that shock acceleration is not very efficient and only 10\% of the available energy eventually gets transferred into
the protons/cosmic-rays ($E_{51}^{\rm amb}$ = 0.77$ E_{51}$ or 0.61$ E_{51}$ in the ``far'' or ``nearby'' case respectively from Eq.\ref{shockedenergy})

The relevant parameters for the two different cases used here are summarized in Table \ref{tab:model}. As shown in
Figure~\ref{fig:mwsed} this conventional hadronic model can  explain very well the whole $\gamma$-ray emission from the GeV to the TeV part of the spectrum in a  way
consistent with all the constraints.

\begin{table}[htbp]
\scriptsize
\begin{center}
\begin{tabular}{|c|c|c|c|c|c|c|}\hline
Case  & $D_{\rm kpc}$ & $n_H$ & $E_{SN}$ & $E_{p,tot}$ & $K_ep$ & E$_{p,max}$ \\
& $[\mathrm{kpc}]$ & $[\mathrm{cm^{-3}}]$ & $[10^{51} \mathrm{erg}]$ & $[10^{50} \mathrm{erg}]$ & 10$^{-4}$ & TeV \\ \hline
Far        & $3.50$ & $0.24$ & $2.0$ & $1.50$ & $4.5 $ & 540\\
Nearby & $2.78$ & $0.30$ & $1.0$ & $0.61$ & $7.0 $ & 340 \\
\hline
\end{tabular}
\caption{\small Parameters used in the spectral energy modeling shown in Figure~\ref{fig:mwsed}.
Spectral indices have been fixed  to $2.3$ for both electrons and protons.} \label{tab:model}
\end{center}
\end{table}

\section{Conclusions}
A 5$\sigma$ detection of GeV $\gamma$-ray emission from Tycho's SNR is reported. The flux above 400~MeV is
(3.5$\pm$1.1$_{stat}\pm$0.7$_{syst}$)$\times$10$^{-9}$ cm$^{-2}$s$^{-1}$ and the  photon index  2.3$\pm$0.2$_{stat}\pm$0.1$_{syst}$ .
The measured Fermi spectrum as well as the available radio,
X-ray and TeV data can be explained by an accelerated proton population which produces $\gamma$-ray photons via $\pi^0$ production and decay. IC emission and bremsstrahlung
can account for only a fraction of the observed $\gamma$-ray flux.

\section{Acknowledgments}
The $Fermi$ LAT Collaboration acknowledges support from a number of agencies and institutes for both development and the operation of the LAT as well as scientific data analysis.
These include
NASA and DOE in the United States, CEA/Irfu and IN2P3/CNRS in France, ASI and INFN in Italy, MEXT, KEK, and JAXA in Japan, and the K.~A.~Wallenberg Foundation, the Swedish
Research Council and
the National Space Board in Sweden. Additional support from INAF in Italy and CNES in France for science analysis during the operations phase is also gratefully
acknowledged.

\end{document}